\documentclass[12pt]{article}
\usepackage{epsfig}

\def\NPB#1#2#3{Nucl. Phys. {\bf B#1}, #2 (19#3)}

\def\PLB#1#2#3{Phys. Lett. {\bf B#1}, #2 (19#3)}

\def\PRD#1#2#3{Phys. Rev. {\bf D#1}, #2 (19#3)}

\def\PRL#1#2#3{Phys. Rev. Lett. {\bf#1}, #2 (19#3)}

\begin{document}


\begin{titlepage}
\noindent
\begin{flushright}
CERN-TH/2000-323\\
\end{flushright}
\begin{center}
  \begin{Large}
    \begin{bf}
      A Constraint on Yukawa-Coupling Unification
      from Lepton-Flavor Violating Processes
    \end{bf}
  \end{Large}
\end{center}
\vspace{0.2cm}
\begin{center}
  J. Sato\\
  \begin{it}
    Research Center for Higher
    Education, Kyushu University,
    Ropponmatsu, Chuo-ku, Fukuoka, 810-8560, Japan\\
  \end{it}
  \vspace{0.5cm}
  K. Tobe\\
  \begin{it}
    CERN -- Theory Division, 
    CH-1211 Geneva 23, Switzerland\\
  \end{it}
  \vspace{0.5cm}
  T. Yanagida\\
  \begin{it}
    Department of Physics, University of Tokyo, Tokyo 113-0033,
    Japan \\ and \\ Research Center for the Early Universe, University of
    Tokyo, Tokyo, 113-0033, Japan
  \end{it}
\end{center}

\begin{abstract}
We present a new constraint on a lepton mixing matrix $V$ from
lepton-flavor violating (LFV) processes in supersymmetric standard
models with massive neutrinos. Here, we assume Yukawa-coupling 
unification $f_{\nu 3}\simeq f_{\rm top}$,
in which $\tau$-neutrino Yukawa coupling $f_{\nu 3}$ is unified into
top-quark Yukawa coupling $f_{\rm top}$ at the unification scale 
$M_*\simeq 3\times 10^{16}$ GeV.
We show that the present experimental bound on $\mu \rightarrow e \gamma$
decay already gives a stringent limit on the lepton
mixing (typically $V_{13}<0.02$ for $V_{23}=1/\sqrt{2}$). Therefore,
many existing neutrino-mass models are strongly constrained.
Future improvement of bounds on LFV processes will provide a 
more significant impact on the models with the Yukawa-coupling unification.
We also stress that a precise measurement of a
neutrino mixing $(V_{MNS})_{e3}$ in future neutrino experiments
would be very important, since the observation of non-zero
$(V_{MNS})_{e3}$, together with negative experimental results 
for the LFV processes, have a robust potential to exclude a large class
of SUSY standard models with the Yukawa-coupling unification.
\end{abstract}
\end{titlepage}

\section{Introduction}
%

It is a remarkable fact that three gauge coupling constants of the
SU(3)$\times$SU(2) $\times$U(1) theory meet at a very high energy scale
$\mu \simeq 3\times 10^{16}$ GeV in the supersymmetric (SUSY) standard
model. The grand unified theory (GUT) is the most manifest candidate to
explain the unification of the three gauge coupling constants. The GUT
was considered as a necessary scheme to maintain the gauge-coupling
unification up to the Planck scale $M_{\rm Planck}\simeq 2\times
10^{18}$ GeV. However, it has been pointed out by Witten some time ago
\cite{witten} that the fundamental scale can be much lower than the
Planck scale in the strongly coupled string (M) theory \cite{harava} and
the unification scale, $\mu \simeq 3\times 10^{16}$ GeV,
is regarded as the cut-off scale $M_*$ of the low-energy effective field
theory. That is, the standard-model gauge interactions are directly
unified with gravity without going through the GUT phase.  This new
interpretation of gauge coupling unification has various
phenomenological merits; in particular, it does not suffer from the
doublet--triplet splitting problem, unlike the SUSY GUT, and it may
provide a natural Peccei--Quinn axion \cite{peccei} to solve the strong
CP problem since the effects of world-sheet instantons are expected to
be suppressed in the strongly coupled string theories \cite{banks}. In
this new unification approach, however, Yukawa coupling constants are
free parameters and hence we need a principle to understand another
success of the SUSY GUT, i.e. $m_{\tau}\simeq m_b$. The Yukawa-coupling
unification is the most well-known principle to explain it. Thus, we
assume the Yukawa-coupling unification for the third family at the
unification (cut-off) scale $M_* \simeq 3\times 10^{16}$ GeV, and
consider that the Yukawa couplings for the first and second families
receive easily large threshold effects from heavy particles at the
cut-off scale, since their tree-level values themselves are small
compared with the Yukawa coupling constants for the third family. This
interesting principle may be extended if there are right-handed
neutrinos.  This is because we may have another ``Yukawa-coupling
unification ($f_{\nu _3}\simeq f_{\rm top}$)'', where $f_{\nu_3}$ is the
largest eigenvalue of the Dirac mass Yukawa coupling for the neutrino.
  
Recently, Super-Kamiokande experiments on atmospheric
neutrinos~\cite{SuperK} have presented very convincing evidence for the
oscillation of $\nu_{\mu}$ to $\nu_{\tau}$ with a mass difference
$\delta m^2 \simeq 10^{-3} - 10^{-2}$~eV$^2$, which implies the largest
mass of the neutrino to be $m_{\nu _3}\simeq 0.03-0.1$ eV, provided
there is a mass
hierarchy $m_{\nu _3}\gg m_{\nu _2}$.  If neutrinos are indeed massive, the
seesaw mechanism~\cite{SeeSaw} is the most natural framework to account
for the smallness of neutrino masses, where the small masses are
low-energy consequences of the presence of superheavy right-handed
neutrinos. The Majorana masses of the right-handed neutrinos are
determined by the Dirac mass term for neutrinos, $m_{\nu_D}$. It is very
interesting that the Yukawa-coupling unification 
($f_{\nu _3}\simeq  f_{\rm top}$) suggests the Majorana mass $M_{R}$ of 
the right-handed neutrino to be $M_R\simeq 2\times 10^{14}$ GeV, which 
is very close to the unification (cut-off) scale $M_*$. 

In the models with the Yukawa-coupling unification $f_{\nu_3}\simeq
f_{\rm top}$, large lepton-flavor violation (LFV) in the charged lepton
sector is expected in the SUSY standard model, since the $\tau$ neutrino
Yukawa coupling is very large and it induces non-negligible mass
splitting among sleptons through radiative corrections \cite{masiero,
HMTYY}. In this letter we perform a detailed analysis of LFV processes
assuming the Yukawa-coupling unification ($f_{\nu_{3}}\simeq
f_{\rm top}$). Throughout this paper we assume that all squarks and sleptons
have a common SUSY-breaking soft mass at the unification scale $M_* \simeq
3\times 10^{16}$ GeV taking the gravity-mediated SUSY breaking model.

We show that the present experimental upper bound on 
$\mu \rightarrow e\gamma$ decay already
gives a stringent constraint on the mixing $V_{13}V_{23}$ 
(typically $V_{13} <0.02$ for $V_{23}=1/\sqrt{2}$).
Since the constraint is very severe, the models with the Yukawa-coupling
unification need an explanation (e.g. symmetry) 
for the smallness of $V_{13}$. In the existing literature, this constraint 
has not been taken into account. Therefore, most of the models with 
the Yukawa-coupling unification should be subject to this new 
constraint. However, we also stress that the above intriguing principle, 
i.e. Yukawa-coupling unification, is not yet ruled out.
Future improvement of the branching ratios for the LFV 
processes~\cite{PSI, MECO, PRISM} will provide a more significant 
impact on the Yukawa-coupling unification models, otherwise they will 
indeed be observed.

Furthermore, in most of the neutrino-mass models proposed so far, the mixing
$V_{13}$ approximately equals the neutrino mixing
$(V_{MNS})_{e3}$~\cite{neutrino_models1,JY,neutrino_models2}. Therefore,
a precise measurement of $(V_{MNS})_{e3}$ in future neutrino
experiments~\cite{joe, neutrino_factory} is very important, since the
observation of non-zero $(V_{MNS})_{e3}$ together with the negative result
of $\mu \rightarrow e\gamma$ decay has a great potential to exclude a
large class of SUSY standard models with the Yukawa-coupling
unification.

\section{SUSY standard model with right-handed neutrinos}
%
First, we briefly review the SUSY standard model with right-handed neutrinos.
Introducing the right-handed neutrinos, we have the following
superpotential in the lepton sector:
\begin{eqnarray}
W &=& \bar{E}_i f_e^{i} L_i H_d + \bar{N}_i f_{\nu}^{ij} L_j H_u
+\frac{1}{2} \bar{N}_i M_{R i} \bar{N}_i + {\rm h.c.},
\end{eqnarray}
where $\bar{E}_i$, $L_i$ and $\bar{N}_i$ are right-handed charged
leptons, lepton doublets and right-handed neutrinos, respectively.  In
this letter, without loss of generality, we take a basis where the
charged-lepton Yukawa couplings ($f_e$) and right-handed neutrino
masses, $M_R$, are diagonalized.

Assuming $M_{R i} \gg m_Z$, we integrate out the right-handed neutrinos
and obtain the following effective superpotential:
\begin{eqnarray}
W &\simeq& \bar{E}_i f_e^{ij} L_j H_d -\frac{1}{2} 
(f_{\nu}^T M_R^{-1} f_{\nu})^{ij} (L_i H_u) (L_j H_u)
 + {\rm h.c.}
\end{eqnarray}
After the electroweak symmetry breaking, neutrinos get tiny Majorana masses:
\begin{eqnarray}
m_{\nu} &=& m^{\rm T}_{\nu_D} M_R^{-1} m_{\nu_D},
\label{neutrino_mass}
\end{eqnarray}
where $m_{\nu_D}=f_\nu \langle H_u\rangle$.
The neutrino mixing matrix $V_{MNS}$ is defined by
\begin{eqnarray}
V^{\rm T}_{MNS} m_{\nu} V_{MNS} &=& {\rm diag.}
(m_{\nu_1},~m_{\nu_2},~m_{\nu_3}),\\
\nu_{F \alpha} &=& (V_{MNS})_{\alpha i} \nu_{M i},
\end{eqnarray}
where $\nu_{F(M)}$ is a flavor-(mass-) eigenstate of
neutrinos.
We also define lepton mixing matrices $V$ and $V_R$, which diagonalize
the neutrino Yukawa matrix as follows:
\begin{eqnarray}
V_R^\dagger f_\nu V &\equiv& f_{\nu}^{\rm diag} = 
{\rm diag.}(f_{\nu 1},f_{\nu 2},f_{\nu 3})
\label{dirac_mixing}\\
V_{MNS}&=&V U.
\end{eqnarray}
Here the matrix $U$ is defined as
\begin{eqnarray}
U^{\rm T} f_{\nu}^{\rm diag} V_R^{\rm T} M_R^{-1} V_R f_{\nu}^{\rm diag} U
= {\rm diag}(\kappa_1,\kappa_2,\kappa_3).
\end{eqnarray}
Note that in general the neutrino mixing matrix $V_{MNS}$ is different
from the lepton mixing matrix $V$. As is well known, the neutrino mixing
matrix $V_{MNS}$ is responsible for the neutrino-oscillation
physics. The lepton mixing matrix $V$, on the other hand, is important
for the LFV processes, as we will see later.  Many models for fermion
masses have been proposed so far, in order to accommodate a large mixing
for atmospheric neutrinos.  It has been pointed out that without
introducing a nontrivial structure in the right-handed Majorana neutrino
mass matrix, the large mixing for neutrinos and small mixings for quarks
are naturally explained by a lopsided structure of the Dirac mass
matrices~\cite{neutrino_models1, JY}. In this case, the lepton mixing
matrix $V$ possesses a large mixing, i.e. $V_{23}\sim$ O(1). As we will
see, the experimental limits on the LFV processes can put a stringent
constraint on the mixing $V$.

We should also note that the neutrino mixing $V_{MNS}$ is approximately
equal to the mixing matrix $V$ in many models~\cite{neutrino_models1,
JY, neutrino_models2}:
\begin{eqnarray}
 V_{MNS}\simeq V.
\label{same}
\end{eqnarray}
In this class of models, neutrino oscillations and LFV phenomena
are correlated. Therefore, neutrino experiments, together with the
LFV searches, will provide a significant constraint on the models.

\section{LFV in SUSY models with right-handed neutrinos}

Let us now discuss the LFV in the SUSY models with right-handed
neutrinos~\cite{masiero,HMTYY}. Non-zero neutrino Yukawa couplings
$f_{\nu}$ generate the LFV in left-handed slepton masses via the
renormalization group (RG) effects, even if a common SUSY-breaking mass
for all scalars is assumed at the unification (cut-off) scale $M_*$. An
approximate solution to the RG equation for masses responsible for LFV
at the weak scale is given by
\begin{eqnarray}
(\Delta m^2_{\tilde{L}})_{ij} &\simeq& -\frac{(6+a_0^2)m_0^2}{16 \pi^2}
(f_\nu^\dagger f_\nu)_{ij} \log\frac{M_*}{M_R},\nonumber \\
&=&-\frac{(6+a_0^2)m_0^2}{16 \pi^2}
V_{ik} V_{jk}^* |f_{\nu k}|^2 \log\frac{M_*}{M_R},~~{\rm for~}i\neq j.
\end{eqnarray}
Here, we have assumed a common SUSY-breaking mass $(m_0)$ for all scalar
bosons and a common A-term $(A_f=a_0 m_0 f_f)$ at the unification scale
($M_*=3\times 10^{16}$ GeV), and we have used Eq.~(\ref{dirac_mixing}).
Note that the LFV masses $(\Delta m^2_{\tilde L})_{ij}$ depend on the
lepton mixing matrix $V$ rather than on the neutrino mixing $V_{MNS}$.
LFV processes $\mu \rightarrow e \gamma$ and $\mu \rightarrow e$ conversion
in nuclei are induced by $(\Delta
m^2_{\tilde{L}})_{21}$ component through the slepton-mediated diagrams.  
Assuming a hierarchical structure of the neutrino Yukawa
couplings $|f_{\nu 3}| \gg |f_{\nu 2}| \gg |f_{\nu 1}|$, 
which is similar to those for the charged-leptons and quarks,
the dominant contribution is given by
\begin{eqnarray}
(\Delta m^2_{\tilde{L}})_{21} &\simeq& -\frac{(6+a_0^2)m^2_0}
{16 \pi^2} V_{23} V_{13}^* |f_{\nu 3}|^2 \log \frac{M_*}{M_R} +\cdots.
\label{LFV_slepton}
\end{eqnarray}
A branching ratio for $\mu \rightarrow e \gamma$
decay is
\begin{eqnarray}
{\rm Br}(\mu \rightarrow e \gamma) \simeq
F \left|\frac{(\Delta m^2_{\tilde{L}})_{21}}{(500~{\rm{GeV}})^2}\right|^2
\tan^2\beta,
\end{eqnarray}
where $F$ is a complicated function of the SUSY parameters.
See Refs.~\cite{HMTYY, Hisano_Nomura} for details.

\begin{figure}
\vspace*{-2.5cm}
\centerline{
\psfig{figure=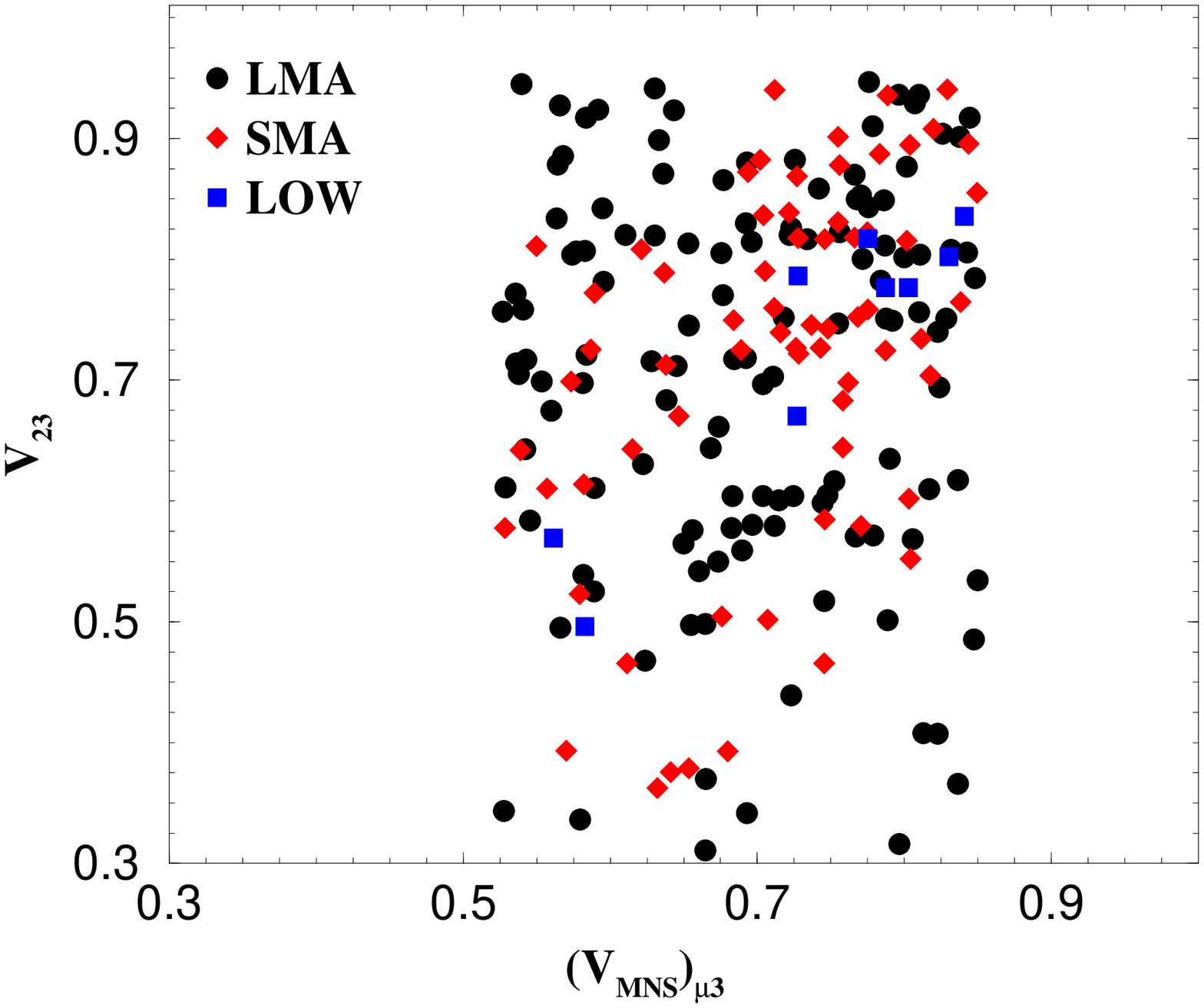,angle=-90,width=0.8\textwidth}
}
\vspace*{-1cm}
\centerline{
\psfig{figure=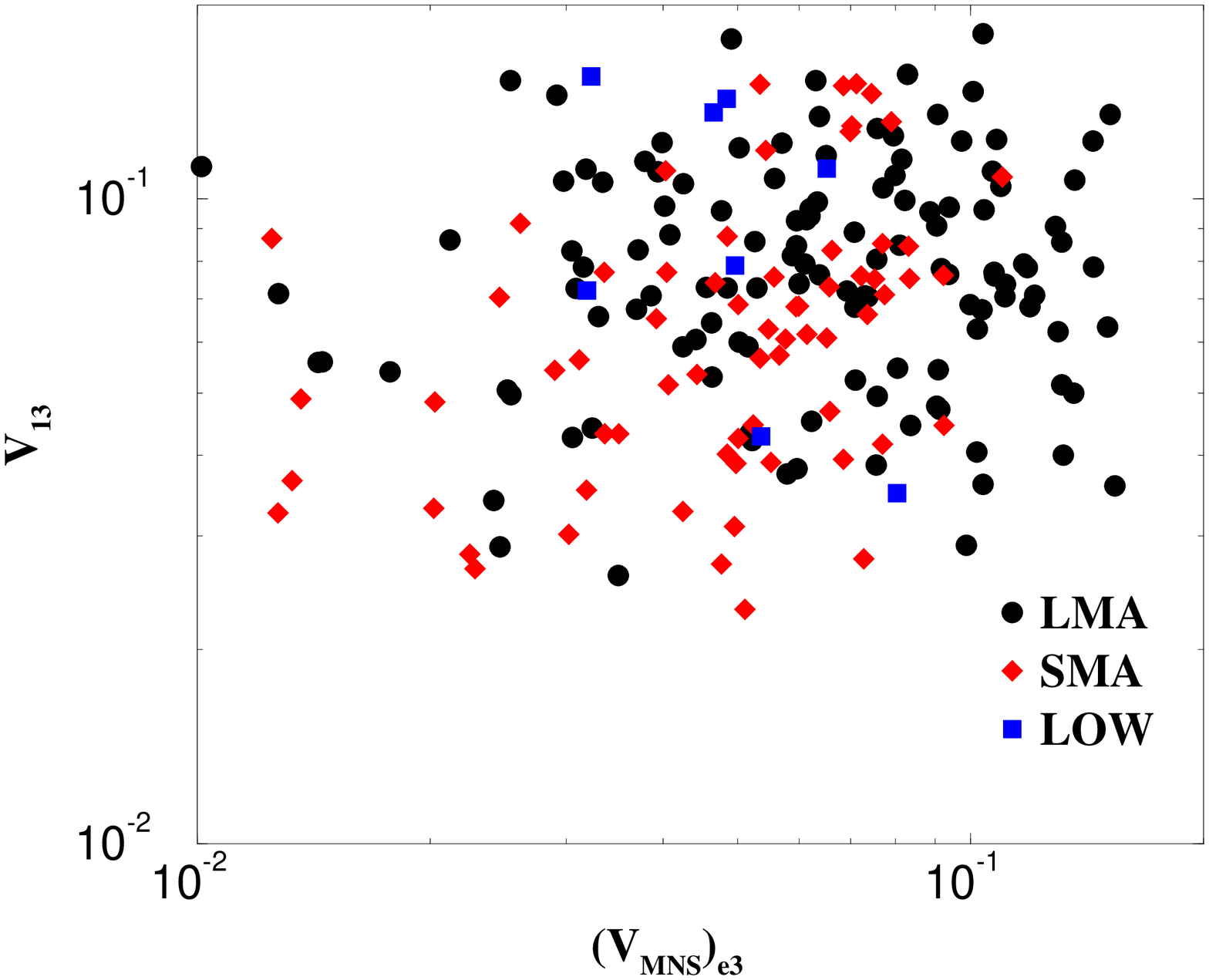,angle=-90,width=0.8\textwidth}
}
\vspace*{0.3cm}
\caption{Predicted values of $(V_{MNS})_{\mu3}~vs.~V_{23}$ and
$(V_{MNS})_{e3}~vs.~V_{13}$ in the neutrino-mass models with 
lopsided Frogatt--Nielsen (FN) U(1) charges in Ref.~\cite{JY}.
Here we considered the FN models where the left-handed lepton doublets
$L_i$ ($i=1$--$3$) have lopsided FN charges, 1, 0, 0, respectively, while
the right-handed charged leptons $\bar{E}_i$ ($i=1$--$3$) and
the right-handed neutrinos $\bar{N}_i$ ($i=1$--$3$)
have the same FN charges, 2, 1, 0, respectively. 
Points denoted by LMA, SMA, and LOW are MSW large mixing angle,
MSW small mixing angle, and LOW solutions for solar neutrinos,
respectively. For more details,
see our forthcoming paper~\cite{JT}. }
\label{v_lepton}
\end{figure}

In the models with the Yukawa-coupling unification, the $\tau$ neutrino
Yukawa coupling is unified into the top-quark Yukawa coupling at
$M_*$. Since the large neutrino Yukawa coupling induces large LFV masses
in the slepton sector~\cite{HMTYY}, the event rates for LFV processes
can be significantly large. As one can see in Eq.~(\ref{LFV_slepton}),
the event rates for $\mu \rightarrow e \gamma$ process 
depend on $V_{23}$ and $V_{13}$
components of the lepton mixing matrix $V$. 
The atmospheric neutrino results indicate that the lepton sector
has a large mixing between the second and third generations.
Especially if the large mixing for atmospheric neutrinos 
originates from the lepton mixing matrix 
$V$, the component $V_{23}$ has a nearly maximal mixing,
$V_{23} \sim 1/\sqrt{2}$, and hence it further enhances the LFV.
For example, the neutrino-mass models with lopsided Froggatt--Nielsen (FN)
U(1) charges in Ref.~\cite{JY} possess a large mixing in $V_{23}$, as
shown in Fig.\ref{v_lepton}. In many of the existing 
models~\cite{neutrino_models1, JY, neutrino_models2},
the almost maximal mixing comes from the lepton mixing $V_{23}$ 
as listed in Table~\ref{list}.
\begin{table}
\begin{center}
\begin{tabular}{|c|c|c|c|}\hline
Models in Refs.~\cite{neutrino_models1, JY, neutrino_models2}
& $V_{23}$ & $V_{13}$ & $ (V_{MNS})_{e3}$ \\
\hline 
Albright {\it et al.} \cite{neutrino_models2} & 0.9 &  0.06 &  0.05\\
Altarelli {\it et al.} \cite{neutrino_models2} & 0.5 & 0.09 &  0.06\\
Bando {\it et al.} \cite{neutrino_models2} 
& $\sim 0.7$ & $\sim 0.1$ & $\sim 0.1$ \\
Hagiwara {\it et al.} \cite{neutrino_models2} & 0.7 & 0.06 & 0.06 \\
Nomura {\it et al.} \cite{neutrino_models2} & 0.7 & $\sim 0.1$ & $\sim 0.1$ \\
Sato {\it et al.} and Buchm\"uller {\it et al.} 
\cite{neutrino_models1, JY}& 0.7 & $\sim 0.05$ & 
$\sim 0.05$ \\
\hline
\end{tabular}
\end{center}
\caption{Typical predicted values for $V_{23}$, $V_{13}$, and $(V_{MNS})_{e3}$
in various models~\cite{neutrino_models1, JY, neutrino_models2, Ellis}.}
\label{list}
\end{table}
Therefore, the searches for LFV can either constrain, or unveil, such a
large lepton mixing, $V_{13}V_{23}$, which cannot be probed by 
neutrino-oscillation physics.

In our analysis, we numerically solve the RG equations and use the complete
formula in Ref.~\cite{HMTYY} for a calculation of the  
$\mu \rightarrow e \gamma$ branching ratio. 
We fix the heaviest neutrino mass to be $5\times 10^{-2}$ eV 
in order to determine the right-handed neutrino mass scale 
($M_R \sim 2\times 10^{14}$ GeV).

\begin{figure}
\centerline{
\psfig{figure=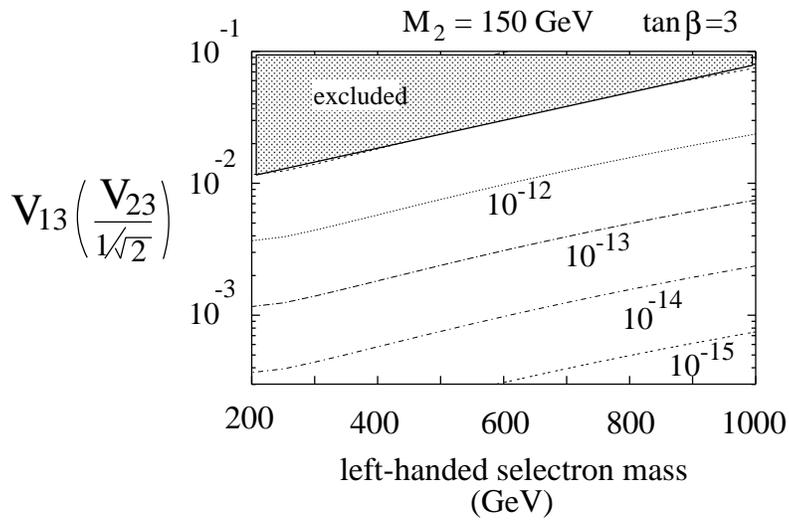,width=0.95\textwidth}
}
\caption{Upper limit on $V_{13} V_{23}$ from the present bound on 
the branching ratio for $\mu \rightarrow e \gamma$ process 
(Br($\mu \rightarrow e \gamma$)$<1.2\times 
10^{-11}$) as a function of the left-handed
selectron mass. Here $V_{23}$ is normalized by $1/\sqrt{2}$, and
we assume the Wino mass  $(M_2)$ to be
150 GeV, and $\tan\beta=3$. 
Numbers on the figure denote Br($\mu \rightarrow e \gamma$).}
\label{exclude}
\end{figure}

In Fig. \ref{exclude}, we present our numerical result, which shows an
upper bound on $V_{13}V_{23}$ from the current limit on the branching ratio
for $\mu\rightarrow e\gamma$, Br$(\mu \rightarrow e \gamma)<1.2\times
10^{-11}$, assuming a Wino mass $(M_2)$ to be 150 GeV, $\tan\beta=3$,
and $f_{\nu 3}=f_{\rm top}$, where $f_{\rm top}$ is
the Yukawa coupling for the top quark at the unification scale $M_*$.
Here, we do not impose the exact bottom-tau Yukawa unification at $M_*$, but
a milder unification ($f_\tau \simeq f_b$ with 20\% deviations)~\cite{polonsky}
is adopted.\footnote{If the exact bottom-tau Yukawa coupling unification
is imposed, we need a large $\tan\beta$ ($\tan\beta \sim
50$)~\cite{smirnov}.  In this case, the constraint on $V_{13} V_{23}$ is more
stringent, since the event rate for $\mu \rightarrow e \gamma$ is nearly
proportional to $\tan^2\beta$.}

In order to obtain a conservative bound on $V_{13}V_{23}$, we have
neglected sub-dominant contributions in the LFV slepton masses in
Eq.~(\ref{LFV_slepton}), which are proportional to $|f_{\nu 2}|^2$ and
$|f_{\nu 1}|^2$.\footnote{If $V_{13}$ is equal to zero, even the
sub-dominant contributions are important. For details, see
Ref.~\cite{Hisano_Nomura}.}  Moreover, since the branching ratio for $\mu
\rightarrow e \gamma$ is approximately proportional to $\tan^2\beta$, we
took a small $\tan\beta$ ($\tan\beta=3$) in Fig.~\ref{exclude}.
Therefore, our result in Fig.~\ref{exclude} should be considered as a
very conservative one.

Our result is also applicable to GUT models with the Yukawa-coupling
unification $f_{\nu 3}=f_{\rm top}$. In the GUT models, the RG running
from the Planck scale to the unification scale also induces the LFV in
slepton masses. Therefore the constraint would be more stringent in
general, unless there are accidental cancellations between the two
contributions below and above the GUT scale, although the GUT
contribution depends on a detail of the model~\cite{GUT}.

As can be seen in Fig.~\ref{exclude}, the current limit on the
Br($\mu\rightarrow e\gamma$) can put a severe bound on $V_{13}V_{23}$;
typically the limit is $V_{13}<0.02$ for $V_{23}=1/\sqrt{2}$ and
$m_{{\tilde e}_L}< 500$ GeV. 
Without any symmetry, such a small value of $V_{13}$ would be very
unnatural. Actually as can been seen in Fig.~\ref{v_lepton} and 
Table~\ref{list}, many of the existing models
are already strongly constrained or
excluded. Furthermore, the future $\mu \rightarrow e \gamma$
experiment~\cite{PSI} with a sensitivity of $10^{-14}$ in the branching
ratio will bring down the limit of $V_{13}$ to $8\times 10^{-4}$ for
$V_{23}=1/\sqrt{2}$ and $m_{{\tilde e}_L}< 500$ GeV, 
and hence it will be  able to
test the models with Yukawa-coupling unification. In addition to the $\mu
\rightarrow e \gamma$ process, $\mu \rightarrow e$ conversion in nuclei
is also important. A ratio between the branching ratio for $\mu
\rightarrow e \gamma$ and the $\mu \rightarrow e$ conversion rate is given
by
\begin{eqnarray}
\frac{{\rm R}(\mu \rightarrow e~{\rm in~Ti~(Al)})}{
{\rm Br}(\mu \rightarrow e \gamma)} \simeq 5~(3)\times 10^{-3},
\end{eqnarray}
in almost the entire parameter space of the models~\cite{HMTYY}. Therefore,
the future MECO experiment~\cite{MECO} for $\mu \rightarrow e$
conversion in Al, with a sensitivity of $10^{-16}$, and a further future
project, with a sensitivity of $10^{-18}$ ({\it e.g.} 
PRISM~\cite{PRISM} for $\mu \rightarrow e$ conversion in Ti) will also
provide a robust probe on the models; otherwise the LFV phenomena will be
observed.\footnote{ A search for another LFV process $\tau \rightarrow
\mu \gamma$, where the event rate depends on the different matrix
elements ($V_{23}$ and $V_{33}$), gives a different information on the
parameters (see Ref.~\cite{JT}).}

We should stress that even if $V_{23}$ is smaller than the maximal
mixing $1/\sqrt{2}$, the constraint does not change much unless it
is extremely small. For example, a factor of 2
smaller value of $V_{23}$ gives only a factor of 2 weaker limit on
$V_{13}$.

Finally, we comment on a connection between neutrino oscillation and
LFV. As we mentioned in the previous section, it is very likely in many
models that the neutrino mixing matrix $V_{MNS}$ equals the lepton
mixing matrix $V$ approximately (see Eq.~(\ref{same})).  Therefore, a
precise measurement of neutrino mixing $(V_{MNS})_{e3}$ in future
neutrino experiments~\cite{joe, neutrino_factory} would be very
important, since the observation of non-zero $(V_{MNS})_{e3}$ could be in
conflict with the constraint from the LFV processes. Therefore, the
neutrino oscillation experiments, together with the LFV searches, have a
strong potential to exclude a large class of SUSY standard models with
the Yukawa-coupling unification.

%

\section*{Acknowledgements}
J.S. is supported in part by a Grant-in-Aid for Scientific
Research of the Ministry of Education, Science and Culture,
\#12047221, \#12740157.
T.Y. is supported in part by the Grant-in-Aid, 
Priority Area ``Supersymmetry and Unified Theory of Elementary
Particles''(\#707).

\end{document}